\documentclass[preprint,review,12pt]{elsarticle}

\usepackage{amssymb}
\usepackage{amsmath}
\usepackage{graphicx}
\usepackage{xcolor}
\usepackage{url}

\renewcommand{\Vec}[1]{\mbox{\boldmath $#1$}}

\usepackage{lineno}

\journal{Computer Physics Communications}

\begin{document}

\begin{frontmatter}

\title{Loading of relativistic Maxwellian-type distributions revisited}
\author{Takayuki Umeda}
\ead{umeda@iic.hokudai.ac.jp}
\address{Information Initiative Center, Hokkaido University, 
Sapporo City, Hokkaido 060-0811, JAPAN}

\begin{abstract}
A simple numerical method for loading of a relativistic Maxwellian-type distribution 
is proposed based on inverse transform sampling. 
The relativistic Maxwellian energy distribution is introduced as 
an alternative to the Maxwell-J\"{u}ttner distribution. 
The cumulative distribution of the shifted-Maxwellian energy distribution 
is approximated by an invertible function. 
Random variates of energy is transformed from uniformly distributed random variables. 
Then, the energy variates are converted to momentum vector variates. 
Numerical tests are presented to show that the present method successfully 
reproduce the relativistic Maxwellian energy distribution. 
\end{abstract}


\end{frontmatter}


\section{Introduction}

Numerical techniques for loading of random variates from a certain distribution 
play an important role 
for initializing particle velocities in particle kinetic simulations 
such as Monte Carlo and particle-in-cell (PIC) simulations. 
There are two methods to generate random variates from a target distribution 
by utilizing random variables from a proposal (source) distribution. 
One is rejection sampling, which accepts a variable from a proposal distribution 
as a variate if  a certain condition due to the target distribution is true, 
otherwise rejects it if false. 
Rejection sampling is inefficient (i.e., the acceptance rate is low) 
if the choice of a proposal distribution is not good. 
Hence, rejection sampling is sometimes unsuitable for high-performance computing.

Alternatively, inverse transform sampling converts random variables from a uniform distribution 
into variates from a target distribution 
by utilizing the inverse cumulative distribution function of the target distribution. 
However, there are many target distributions that are not integrable analytically. 
There are also many cumulative distributions that are not invertible analytically.

For non-relativistic particles, 
the Maxwell-Boltzmann or Maxwellian velocity distribution is commonly used 
as an initial velocity distribution. 
There is a classic inverse transform method for the Maxwellian distribution in two dimensions, 
which is well known as the Box-Muller transform \citep{Box_1958}. 
It is easy to extend the Box-Muller transform for 
isotropic Maxwellian velocity distributions 
to anisotropic and/or drifting (shifted) non-relativistic velocity distributions.

For relativistic particles, 
the Maxwell-J\"{u}ttner distribution \citep{Juttner} is commonly used 
as an initial momentum distribution. 
Because of its difficulty and importance in numerical simulations, 
a numerical method for generating random variates from the Maxwell-J\"{u}ttner distribution 
is an issue for a half century (see Ref. \citep{Zenitani_2022} and references therein). 
Rejection sampling is widely used for the Maxwell-J\"{u}ttner distribution  \citep{Pozdnyakov_1983,Canfield_1987,Swisdak_2013,Schnittman_2013,Zenitani_2015,Zenitani_2022,Zenitani_2024}.
To use inverse transform sampling, 
a numerical table for the cumulative distribution of the Maxwell-J\"{u}ttner distribution 
is necessary \citep{Melzani_2013}, because it is difficult to obtain an analytic expression 
for the cumulative distribution. 
However, the resolution of the table and the procedure of numerical interpolation affect 
the accuracy of the inverse cumulative distribution. 


In the present study, a simple numerical method for loading of a relativistic and shifted 
Maxwellian-type distribution 
is proposed based on inverse transform sampling. 
In Sec. 2, the relativistic Maxwellian energy distribution is introduced 
as an alternative to the Maxwell-J\"{u}ttner distribution. 
Then a numerical method for loading of the relativistic shifted-Maxwellian energy distribution 
is presented. 
In Sec. 3, a comparison is made between the relativistic Maxwellian energy distribution 
and the Maxwell-J\"{u}ttner distribution. 
Properties (mean velocity, mean momentum, thermal energy and kinetic energy) 
of the two distributions are also given. 
Finally in Sec. 4, the summary of the present study is presented.

\section{Maxwellian Energy Distribution}

\subsection{Non-relativistic and isotropic Maxwellian}

The non-relativistic and isotropic Maxwellian velocity distribution is given as 
\begin{equation}
\label{eq:non-rela-maxwellian}
f\left(\Vec{v}\right) = \left( \sqrt{\frac{m}{2\pi T}} \right)^3 \exp\left(-\frac{m\left|\Vec{v}\right|^2}{2T}\right),
\end{equation}
where
\begin{equation*}
\label{eq:non-rela-maxwellian-integral}
\int_{-\infty}^{\infty}\int_{-\infty}^{\infty}\int_{-\infty}^{\infty} f\left(\Vec{v}\right) {\rm d}^3\Vec{v} = 1.
\end{equation*}

For an isotropic velocity distribution, 
the velocity vector $\Vec{v}$ in Cartesian coordinates
is converted to spherical coordinates as follows, 
\begin{equation*}
{\rm d}^3\Vec{v} = v^2\sin\theta {\rm d}v{\rm d}\theta{\rm d}\phi = 4\pi v^2 {\rm d}v.
\end{equation*}
Then, equation (\ref{eq:non-rela-maxwellian}) is written in terms of the scalar velocity 
$v \equiv \left|\Vec{v}\right|$ as 
\begin{equation}
\label{eq:non-rela-maxwellian-v}
f_v\left(v\right) = 
\sqrt{\frac{2}{\pi}} \left( \sqrt{\frac{m}{T}} \right)^3 v^2 \exp\left(-\frac{mv^2}{2T}\right), 
\end{equation}
where
\begin{equation*}
\int_{0}^{\infty} f_v\left(v\right) {\rm d}v = 1. 
\end{equation*}
Wolfram Mathematica \citep{Mathematica} is used for confirmation of this integral.

Alternatively, equation (\ref{eq:non-rela-maxwellian-v}) is also rewritten in terms of the kinetic energy 
${\cal K} \equiv mv^2/2$ as \citep{Livadiotis_2016} 
\begin{equation}
\label{eq:non-rela-maxwellian-e}
f_{\cal K} \left({\cal K}\right) = \frac{2}{\sqrt{\pi}T} \sqrt{\frac{\cal K}{T}} \exp\left(-\frac{\cal K}{T}\right), 
\end{equation}
where
\begin{equation*}
\int_{0}^{\infty} f_{\cal K}\left({\cal K}\right) {\rm d}{\cal K} = 1. 
\end{equation*}
This integral is confirmed with Mathematica.
Note that the following property is used, 
\begin{equation*}
{\rm d}{\cal K} = mv{\rm d}v, \ \ \ 
\sqrt{\frac{2\cal K}{m^3}} {\rm d}{\cal K} = v^2 {\rm d}v. 
\end{equation*}
Equation (\ref{eq:non-rela-maxwellian-e}) is known as the Maxwellian ``energy'' distribution.

\subsection{Relativistic and isotropic Maxwellian}

The relativistic and isotropic Maxwellian energy distribution is given by replacing 
${\cal K}$ in Eq. (\ref{eq:non-rela-maxwellian-e}) with $mc^2(\gamma-1)$ as 
\begin{equation}
\label{eq:rela-maxwellian-g}
f_{\gamma} \left(\gamma \right) = \frac{2}{\sqrt{\pi}} \frac{mc^2}{T}\sqrt{\frac{mc^2}{T}(\gamma-1)} 
\exp\left\{-\frac{mc^2}{T}(\gamma-1)\right\} 
\end{equation}
with $\gamma$ being the relativistic Lorentz factor, 
where 
\begin{equation*}
\int_{1}^{\infty} f_{\gamma}\left(\gamma\right) {\rm d}\gamma = 1. 
\end{equation*}
The thermal energy is given by performing the following integral, 
\begin{equation}
\int_{1}^{\infty} mc^2(\gamma-1)f_{\gamma}\left(\gamma\right) {\rm d}\gamma = \frac{3}{2}T. 
\end{equation}
These integrals are confirmed with Mathematica. 

Equation (\ref{eq:rela-maxwellian-g}) is written in terms of the scalar momentum 
$u \equiv \gamma v$ as 
\begin{equation}
\label{eq:rela-maxwellian-u}
f_u \left(u\right) = \frac{2}{\sqrt{\pi}} 
\left(\sqrt{\frac{m}{T}}\right)^3
\frac{u^2}{\sqrt{1+\frac{u^2}{c^2}}\sqrt{\sqrt{1+\frac{u^2}{c^2}}+1} }
\exp\left\{-\frac{mc^2}{T}\left(\sqrt{1+\frac{u^2}{c^2}}-1\right)\right\}, 
\end{equation}
where
\begin{equation*}
\int_{0}^{\infty} f_u\left(u\right) {\rm d}u = 1.
\end{equation*}
This integral is confirmed with Mathematica.
Note that the following property is used, 
\begin{equation*}
\gamma = \sqrt{1+\frac{u^2}{c^2}}, \ \ \ 
c^2\gamma{\rm d}{\cal \gamma} = u{\rm d}u. 
\end{equation*}

Alternatively, equation (\ref{eq:rela-maxwellian-g}) is also rewritten in terms of the momentum vector 
$\Vec{u}$ as 
\begin{equation}
\label{eq:rela-maxwellian}
f\left(\Vec{u}\right) = \frac{1}{2} 
\left(\sqrt{\frac{m}{\pi T}}\right)^3
\frac{1}{
\gamma\left(\Vec{u}\right)
\sqrt{
\gamma\left(\Vec{u}\right)
+1} }
\exp\left[-\frac{mc^2}{T}\left\{
\gamma\left(\Vec{u}\right)
-1\right\}\right], 
\end{equation}
where
\begin{equation*}
\int_{-\infty}^{\infty}\int_{-\infty}^{\infty}\int_{-\infty}^{\infty} f\left(\Vec{u}\right) {\rm d}^3\Vec{u} = 1.
\end{equation*}
This integral is confirmed by using a numerical integration. 
Note that the following property is used, 
\begin{equation}
\label{eq:convert}
{\rm d}^3\Vec{u} = 
u^2\sin\theta{\rm d}u{\rm d}\theta{\rm d}\phi = 
c^3\gamma\sqrt{\gamma^2-1}\sin\theta {\rm d}{\cal \gamma}{\rm d}\theta{\rm d}\phi 
\end{equation}
with 
\begin{equation*}
\int_0^\pi \sin\theta {\rm d}\theta=2, \ \ \ \int_0^{2\pi}{\rm d}\phi=2\pi. 
\end{equation*}

\subsection{Relativistic shifted-Maxwellian}

A relativistic shifted-Maxwellian energy distribution 
in terms of the momentum vector is given by replacing $\gamma$ 
in Eq. (\ref{eq:rela-maxwellian}) with the boosted Lorentz factor $\gamma_B$ 
and performing renormalization as 
\begin{equation}
\label{eq:rela-d-maxwellian}
f\left(\Vec{u}\right) = \frac{1}{2\gamma_D} 
\left(\sqrt{\frac{m}{\pi \gamma_D T}}\right)^3
\frac{1}{
\gamma_B\left(\Vec{u}\right)
\sqrt{
\gamma_B\left(\Vec{u}\right)
+1} }
\exp\left[-\frac{mc^2}{\gamma_D T}\left\{
\gamma_B\left(\Vec{u}\right)
-1\right\}\right], 
\end{equation}
where 
\begin{equation}
\label{eq:gmb}
\gamma_B = \gamma_D \left( \gamma - \frac{\Vec{v}_D\cdot\Vec{u}}{c^2} \right)
\end{equation}
with the drift velocity vector $\Vec{v}_D$ and its Lorentz factor 
\begin{equation}
\gamma_D = \frac{c^2}{\sqrt{c^2-\left|\Vec{v}_D\right|^2}}.
\end{equation}
The following integral is confirmed by using a numerical integration, 
\begin{equation*}
\int_{-\infty}^{\infty} \int_{-\infty}^{\infty} \int_{-\infty}^{\infty} f\left(\Vec{u}\right)
{\rm d}^3\Vec{u} =1.
\end{equation*}

The transformation from 
the Lorentz-factor-spherical coordinate to the momentum-vector-Cartesian coordinate 
is performed with following properties, 
\begin{subequations}
\label{eq:rela-d-maxwellian-rand}
\begin{equation}
u_{x} = \gamma_D\left( c\sqrt{\gamma_{B}^2-1} \cos\theta + \gamma_{B} v_D \right),
\end{equation}
\begin{equation}
u_{y} = c\sqrt{\gamma_{B}^2-1} \sin\theta \cos\phi,
\end{equation}
\begin{equation}
u_{z} = c\sqrt{\gamma_{B}^2-1} \sin\theta \sin\phi. 
\end{equation}
\end{subequations}
Here, the momentum vector coordinate is taken so that 
the $u_x$ axis is parallel to the drift velocity vector $\Vec{v}_D$. 
Then, the following property is obtained by using the Jacobian determinant 
(see \ref{sec:determinant} for details), 
\begin{equation}
\label{eq:convert_gb}
{\rm d}^3\Vec{u} = c^2\gamma_D \left(
c\gamma_B+v_D\sqrt{\gamma_B^2-1}\cos\theta
\right) 
\sqrt{\gamma_B^2-1}\sin\theta
{\rm d}\gamma_B{\rm d}\theta{\rm d}\phi. 
\end{equation}
Hence, the relativistic shifted-Maxwellian energy distribution 
in terms of the boosted Lorentz factor and the polar angle is given as follows, 
\begin{equation}
\label{eq:rela-d-maxwellian-gt}
f_{\gamma_B,\theta}\left( \gamma_B,\theta \right) = \frac{1}{\sqrt{\pi}}
\left(\sqrt{\frac{mc^2}{\gamma_DT}}\right)^3
\left(
1+\frac{v_D}{c\gamma_B}\sqrt{\gamma_B^2-1}\cos\theta
\right)
\sqrt{\gamma_B-1}\sin\theta
\exp\left\{-\frac{mc^2}{\gamma_DT}\left(
\gamma_B
-1\right)\right\},
\end{equation}
where
\begin{equation*}
\int_{1}^{\infty} \int_{0}^{\pi} 
f_{\gamma_B,\theta}\left(\gamma_B,\theta\right) {\rm d}\theta {\rm d}\gamma_B= 1. 
\end{equation*}
This integral is confirmed by using Mathematica. 
By integrating Eq. (\ref{eq:rela-d-maxwellian-gt}) over $\theta$, the following distribution 
in terms of the boosted Lorentz factor 
is obtained, which is identical to Eq. (\ref{eq:rela-maxwellian-g}), 
\begin{equation}
\label{eq:rela-d-maxwellian-g}
f_{\gamma_B}\left(\gamma_B\right) = \int_{0}^{\pi} 
f_{\gamma_B,\theta}\left(\gamma_B,\theta\right) {\rm d}\theta
= \frac{2}{\sqrt{\pi}}
\left(\sqrt{\frac{mc^2}{\gamma_DT}}\right)^3
\sqrt{\gamma_B-1}
\exp\left\{-\frac{mc^2}{\gamma_DT}\left(
\gamma_B
-1\right)\right\}.
\end{equation}

\subsection{Inverse transform sampling}

With the normalized (dimensionless) energy ${\cal E} = \frac{mc^2}{\gamma_DT}(\gamma_B-1)$, 
equation (\ref{eq:rela-d-maxwellian-g}) is rewritten as 
\begin{equation}
\label{eq:rela-d-maxwellian-e}
f_{\cal E} \left({\cal E} \right) = \frac{2}{\sqrt{\pi}} \sqrt{\cal E} 
\exp\left\{-\cal E\right\}, 
\end{equation}
where 
\begin{equation*}
\int_{0}^{\infty} f_{\cal E} \left({\cal E} \right) {\rm d}{\cal E} = 1. 
\end{equation*}
This integral is confirmed with Mathematica.
The cumulative distribution of Eq. (\ref{eq:rela-d-maxwellian-e}) is given as 
\begin{equation}
F_{\cal E}(x) = \int_0^x f_{\cal E} \left({\cal E} \right) {\rm d}{\cal E} 
= 1 - \frac{2}{\sqrt{\pi}} \Gamma\left(\frac{3}{2},x\right)
\label{eq:rela-d-maxwellian-e-cumu}
={\rm erf}\left(\sqrt{x}\right) - \frac{2}{\sqrt{\pi}}\sqrt{x}e^{-x}.  
\end{equation}
This integral is confirmed with Mathematica. 
Here, $\Gamma\left(n,x\right)$ is the upper incomplete Gamma function, 
\begin{equation*}
\Gamma\left(\frac{3}{2},x\right) = \frac{1}{2}\Gamma\left(\frac{1}{2},x\right) + \sqrt{x}e^{-x} 
= \frac{\sqrt{\pi}}{2}\left\{1-{\rm erf}\left(\sqrt{x}\right)\right\} + \sqrt{x}e^{-x}. 
\end{equation*}

Since there is no analytic inverse function of the cumulative distribution
in Eq. (\ref{eq:rela-d-maxwellian-e-cumu}),  
the following approximation is utilized \cite{Soranzo_2014,Umeda_2024}, 
\begin{equation}
\label{eq:rela-d-maxwellian-e-app}
F_{\cal E}(x) \approx \left\{ 1-\exp\left(-\frac{ax+bx^2}{1+cx+dx^2}\right) \right\}^\frac{3}{2} 
\equiv F_{\rm app}(x).
\end{equation}
It is noted that the coefficients $a$, $b$, $c$ and $d$ must be constant 
to make Eq. (\ref{eq:rela-d-maxwellian-e-app}) invertible. 
The Taylor series expansion of Eqs. (\ref{eq:rela-d-maxwellian-e-cumu}) and (\ref{eq:rela-d-maxwellian-e-app})
around $x=0$ 
are derived with Mathematica as follows, 
\begin{equation*}
F_{\cal E}(x) \approx \frac{4}{3\sqrt{\pi}}x^{\frac{3}{2}} + O\left(x^{\frac{5}{2}}\right), 
\end{equation*}
\begin{equation*}
F_{\rm app}(x) \approx \left(ax\right)^\frac{3}{2} + O\left(x^{\frac{5}{2}}\right). 
\end{equation*}
Hence, the coefficient $a$ is obtained as 
\begin{equation*}
a = \left(\frac{16}{9\pi} \right)^\frac{1}{3}. 
\end{equation*}
To obtain the other coefficients $b$, $c$ and $d$, 
a brute-force search is performed to minimize the root mean square of numerical errors between 
Eqs. (\ref{eq:rela-d-maxwellian-e-cumu}) and (\ref{eq:rela-d-maxwellian-e-app}) for $0 < x \le 8$, 
which gives  
$b=-3.12562\times10^{-2}$, $c=-5.15921\times10^{-2}$ and $d=8.84448\times10^{-4}$. 

\begin{figure}[t]
\center
\includegraphics[width=0.5\textwidth,bb=0 0 640 520]{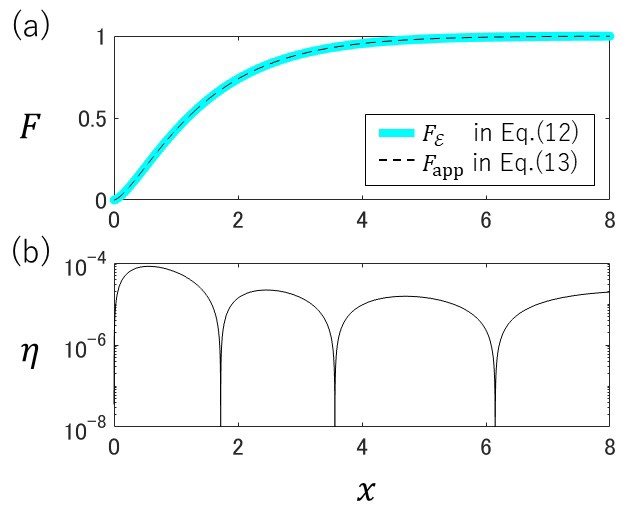}
\caption{
Comparison between the cumulative distribution $F_{\cal E}$ in Eq. (\ref{eq:rela-d-maxwellian-e-cumu}) 
and its approximation $F_{\rm app}$ in Eq. (\ref{eq:rela-d-maxwellian-e-app}). 
(a) The profiles of $F_{\cal E}$ and  $F_{\rm app}$ for $0 < x \le 8$. 
The thick line shows $F_{\cal E}$ and the dashed line shows $F_{\rm app}$. 
(b) The relative error $\eta$ between  $F_{\cal E}$ and $F_{\rm app}$. 
}
\label{fig:1}
\end{figure}

Figure \ref{fig:1} shows the comparison between the cumulative distribution 
in Eq. (\ref{eq:rela-d-maxwellian-e-cumu}) 
and its approximation in Eq. (\ref{eq:rela-d-maxwellian-e-app}) 
with the coefficients listed above. 
In Panel (a), spatial profiles of these functions are shown. 
In Panel (b), the relative error between 
Eqs. (\ref{eq:rela-d-maxwellian-e-cumu}) and (\ref{eq:rela-d-maxwellian-e-app}), 
i.e., $\eta = \left|(F_{\cal E}-F_{\rm app})/F_{\cal E}\right|$, is shown. 
It is found that the relative error of the approximated cumulative distribution 
is less than $10^{-4}$. 

The inverse function of Eq. (\ref{eq:rela-d-maxwellian-e-app}) is given as 
\begin{equation*}
{\cal Y} = \log\left(1-y^\frac{2}{3}+\varepsilon \right),
\end{equation*}
\begin{equation}
\label{eq:rela-d-maxwellian-e-cumu-app-inv}
F_{\rm app}^{-1}\left( y \right) = \frac{\sqrt{\left\{a+c{\cal Y}(y)\right\}^2-4{\cal Y}(y)\left\{b+d{\cal Y}(y)\right\}}
-\left\{a+c{\cal Y}(y)\right\}}{2\left\{b+d{\cal Y}(y)\right\}},
\end{equation}
for $0\le x < 1$. 
Here, $\varepsilon$ denotes a small variable (e.g., $2^{-126}$) to avoid taking zero 
as an argument of the logarithm function.

Random variates from Maxwellian energy distribution are 
generated with random variables from uniform distribution, $0 \le R_n^{(1)}<1$, 
as follows, 
\begin{equation}
\label{eq:rela-d-maxwellian-e-rand}
{\cal E}_n = F_{\rm app}^{-1}\left( R_n^{(1)} R_{\rm ul} \right). 
\end{equation}
where $R_{\rm ul}$  denotes the upper limit for 
the argument of the inverse function 
in Eq. (\ref{eq:rela-d-maxwellian-e-cumu-app-inv}), 
which is introduced to avoid taking a negative number as an operand of 
square root in Eq. (\ref{eq:rela-d-maxwellian-e-cumu-app-inv}). 
With the coefficients listed above, the upper limit is given as $R_{\rm ul} \approx 0.999997546$.

\begin{figure}[t]
\center
\includegraphics[width=0.5\textwidth,bb=0 0 640 580]{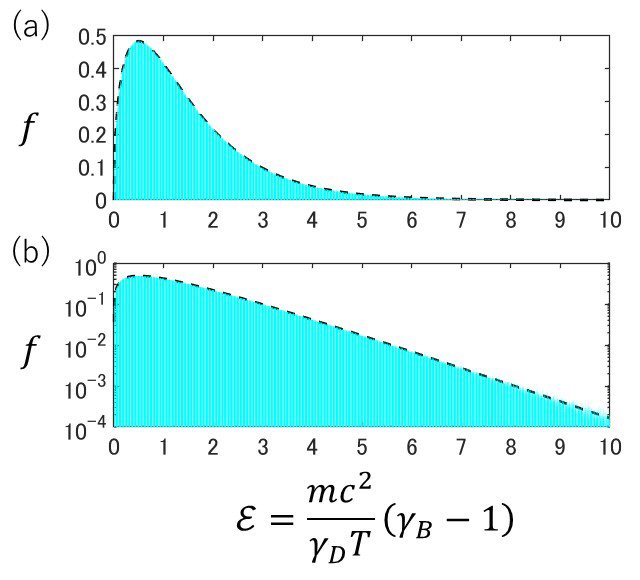}
\caption{
Comparison between the Maxwellian energy distribution 
in Eq. (\ref{eq:rela-d-maxwellian-e}) and a histogram of numerically generated random variates. 
The dashed lines show the profile of Eq. (\ref{eq:rela-d-maxwellian-e}) 
as a function of the normalized energy ${\cal E}=mc^2(\gamma_B-1)/(\gamma_DT)$ 
in (a) a linear scale and (b) a logarithmic scale. 
The bars show a histogram of random variates generated 
by using Eq. (\ref{eq:rela-d-maxwellian-e-rand}) 
with $N_{\rm s}=10^8$ samples and $N_{\rm bin}=10^4$ bins. 
}
\label{fig:2}
\end{figure}

Figure \ref{fig:2} shows the comparison between the Maxwellian energy distribution 
in Eq. (\ref{eq:rela-d-maxwellian-e}) and a histogram of numerically generated random variates. 
The dashed lines show the profile of Eq. (\ref{eq:rela-d-maxwellian-e}) 
as a function of the normalized energy ${\cal E}=mc^2(\gamma_B-1)/(\gamma_DT)$ 
in (a) a linear scale and (b) a logarithmic scale. 
The bars show a histogram of random variates generated 
by using Eq. (\ref{eq:rela-d-maxwellian-e-rand}). 
The histogram is created with $N_{\rm s}=10^8$ samples and $N_{\rm bin}=10^4$ bins. 
The result shows an excellent agreement between the analytic distribution and the histogram. 


In general, variates for energy (or scalar momentum) in spherical coordinates 
are converted to variates for momentum vector 
with other two sets of random variables from uniform distribution, 
$0 \le R_n^{(2)} < 1$ and $0 \le R_n^{(3)} < 1$,  
via 
\begin{subequations}
\label{eq:theta-phi}
\begin{equation}
\label{eq:theta}
\theta_n = \cos^{-1}\left(2R_n^{(2)} -1 \right), 
\end{equation}
\begin{equation}
\label{eq:phi}
\phi_n = 2\pi R_n^{(3)}. 
\end{equation}
\end{subequations}
Equation (\ref{eq:theta}) is rewritten as 
the following well-known identities, 
\begin{equation*}
\cos\theta_n = \cos\left[\cos^{-1} \left(2R_n^{(2)}-1\right)\right]=2R_n^{(2)}-1,
\end{equation*}
\begin{equation*}
\sin\theta_n = \sin\left[\cos^{-1} \left(2R_n^{(2)}-1\right)\right]=\sqrt{1-\left(2R_n^{(2)}-1\right)^2} 
= 2\sqrt{R_n^{(2)}\left(1-R_n^{(2)}\right)}. 
\end{equation*}
For a general relativistic distribution, however, 
it is not is able to transform from the boosted-Lorentz-factor coordinate (i.e., $f_{\gamma_B}$) 
to the momentum-vector coordinate (i.e., $f$) 
by using Eq. (\ref{eq:theta-phi}). 
As shown below,  
the distribution obtained by the coordinate transformation with Eq. (\ref{eq:theta-phi})
is not an equilibrium distribution (i.e., a function of $\gamma_B$) 
unlike Eq. (\ref{eq:rela-d-maxwellian}) (see \ref{sec:determinant} for details), 
\begin{equation*}
f \left(\Vec{u} \right) = \frac{4\pi}{\gamma\left(\Vec{u} \right)\sqrt{ \gamma_B^2\left(\Vec{u} \right) -1}}
f_{\gamma_B}\left\{ \gamma_B\left(\Vec{u} \right) \right\}. 
\end{equation*}

A general relativistic distribution in terms of the momentum vector ($f$)  
is converted to a relativistic distribution in terms of both boosted Lorentz factor 
and polar angle ($f_{\gamma_B,\theta}$) with Eq. (\ref{eq:convert_gb}) 
as follows, 
\begin{equation*}
f_{\gamma_B,\theta}\left( \gamma_B,\theta \right) = 
\frac{c^2}{4\pi} \sqrt{\gamma_B^2-1}\sin\theta
\left( c\gamma_B+v_D\sqrt{\gamma_B^2-1}\cos\theta \right)
f\left(\Vec{u} \right).
\end{equation*}
Since the relativistic distribution is expressed as a function of the polar angle 
as well as the boosted Lorentz factor, 
it is necessary to take the $v_D(\ne 0)$ term into account, 
which is excluded in Eq. (\ref{eq:theta}). 
The normalized cumulative distribution of Eq. (\ref{eq:rela-d-maxwellian-gt}) 
in terms of the polar angle is given as 
\begin{equation}
\label{eq:rela-d-maxwellian-gt-cum}
F_{\gamma_B,\theta}\left( x \right) =
\frac{1}{2}\left(1-\cos x\right)+\frac{v_D}{2c}\sqrt{1-\frac{1}{\gamma_B^2}}\left(1-\cos^2 x\right).
\end{equation}
The inverse function of this cumulative distribution is given as 
\begin{equation}
\label{eq:rela-d-maxwellian-gt-cum-inv}
F_{\gamma_B,\theta}^{-1} \left( y \right) = \cos^{-1}\left[
\frac{\sqrt{1+\frac{v_D^2}{c^2}\left(1-\frac{1}{\gamma_B^2}\right)
+\frac{2v_D}{c}\sqrt{1-\frac{1}{\gamma_B^2}}(1-2y)}-1}{\frac{v_D}{c}\sqrt{1-\frac{1}{\gamma_B^2}}}
\right].
\end{equation}

To summarize, 
three sets of random variables from uniform distribution, $R_n^{(1)}$, $R_n^{(2)}$ and $R_n^{(3)}$, 
are converted to variates of boosted Lorentz factor, polar angle, and azimuth angle as follows, 
\begin{subequations}
\label{eq:summary}
\begin{equation}
\gamma_{B,n} = 1+\frac{\gamma_DT}{mc^2} F_{\rm app}^{-1}\left(R_n^{(1)}R_{\rm ul}\right), 
\end{equation}
\begin{equation}
\theta_n = 
\left\{\begin{array}{ll}
F_{\gamma_B,\theta}^{-1} \left( R_n^{(2)} \right) & \left(v_D \ne 0 \right), \\
\cos^{-1}\left(2R_n^{(2)}-1\right) & \left(v_D = 0 \right), \\
\end{array}
\right.
\end{equation}
\begin{equation}
\phi_n = 2\pi R_n^{(3)},
\end{equation}
\end{subequations}
by using the inverse functions in Eqs. (\ref{eq:rela-d-maxwellian-e-cumu-app-inv}) 
and (\ref{eq:rela-d-maxwellian-gt-cum-inv}). 
Then, these variates are transformed into variates of momentum vector 
by using Eq. (\ref{eq:rela-d-maxwellian-rand}). 
With these procedures, 
the relativistic shifted-Maxwellian energy distribution in terms of the momentum vector 
in Eq. (\ref{eq:rela-d-maxwellian}) is obtained. 

With the following property \citep{Ueno_2021}, 
\begin{equation*}
{\rm d}^3\Vec{u} = \gamma^5 {\rm d}^3\Vec{v}, 
\end{equation*}
the momentum distribution in Eq. (\ref{eq:rela-d-maxwellian}) is rewritten in terms of the velocity vector 
as follows, 
\begin{equation}
\label{eq:rela-d-maxwellian-v}
f_v\left(\Vec{v}\right) = \frac{1}{2\gamma_D} 
\left(\sqrt{\frac{m}{\pi \gamma_DT}}\right)^3
\frac{\gamma^5\left(\Vec{v}\right)}{
\gamma_B\left(\Vec{v}\right)
\sqrt{
\gamma_B\left(\Vec{v}\right)
+1} }
\exp\left[-\frac{mc^2}{\gamma_DT}\left\{
\gamma_B\left(\Vec{v}\right)
-1\right\}\right], 
\end{equation}
where 
\begin{equation*}
\int_{-\infty}^{\infty} \int_{-\infty}^{\infty} \int_{-\infty}^{\infty} f\left(\Vec{v}\right)
{\rm d}^3\Vec{v} =1.
\end{equation*}
This integral is confirmed by using a numerical integration.

\begin{figure}[t]
\center
\includegraphics[width=1.0\textwidth,bb=0 0 1280 520]{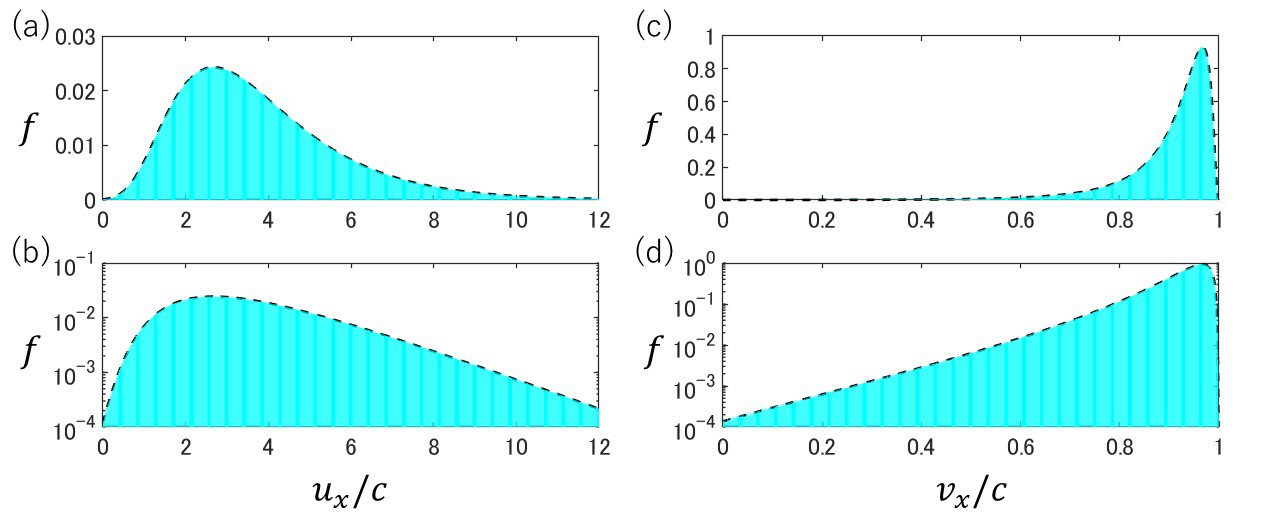}
\caption{
Comparison between the momentum/velocity distribution 
based on the Maxwellian energy distribution 
and a histogram of numerically generated random variates. 
The dashed lines in the left panels show the reduced momentum distribution in the $u_x$ direction
with $v_D/c=0.9$ and $mc^2/T=6.25$ in (a) a linear scale and (b) a logarithmic scale. 
The dashed lines in the right panels show the corresponding reduced velocity distribution 
in the $v_x$ direction in (c) a linear scale and (d) a logarithmic scale. 
The bars show a histogram of random variates generated 
by using Eqs. (\ref{eq:rela-d-maxwellian-rand}) and (\ref{eq:summary})
with $N_{\rm s}=10^8$ samples and $N_{\rm bin}=10^4$ bins. 
}
\label{fig:3}
\end{figure}

Figure \ref{fig:3} shows the comparison between the momentum/velocity distribution  
and a histogram of numerically generated random variates. 
The dashed lines in the left panels show the reduced momentum distribution in terms of $u_x$  
with $v_D/c=0.9$ and $mc^2/T=6.25$. 
The dashed lines in the right panels show the corresponding reduced velocity distribution in terms of $v_x$. 
Since it is not easy to analytically integrate the distributions in Eqs. (\ref{eq:rela-d-maxwellian}) 
and (\ref{eq:rela-d-maxwellian-v}) over the $u_y-u_z$ space and the $v_y-v_z$ space  
(i.e., the momentum/velocity space perpendicular to the drift velocity vector) respectively, 
a numerical integration is performed to obtain the reduced momentum and velocity distributions 
in the $u_x$ and $v_x$ directions. 
The bars show a histogram of random variates generated 
by using Eqs. (\ref{eq:rela-d-maxwellian-rand}) and (\ref{eq:summary}). 
The histogram is created with $N_{\rm s}=10^8$ samples and $N_{\rm bin}=10^4$ bins. 
The result shows an excellent agreement between the analytic distribution and the histogram.

\section{Comparison against Maxwell-J\"{u}ttner Distribution}

The Maxwell-J\"{u}ttner distribution in terms of the momentum vector $\Vec{u}$ is 
given as 
\begin{equation}
\label{eq:juttner}
g\left(\Vec{u}\right) = \frac{m}{4\pi c T K_2\left(\frac{mc^2}{T}\right)}
\exp\left(-\frac{mc^2}{T}\sqrt{1+\frac{\left|\Vec{u}\right|^2}{c^2}}\right),
\end{equation}
where $K_n$ denotes the modified Bessel function of the second kind of order $n$. 
For an isotropic distribution, 
this distribution is alternatively written by utilizing Eq. (\ref{eq:convert})
in terms of the scalar momentum $u$ as 
\begin{equation}
\label{eq:juttner-u}
g_u\left(u\right) = \frac{mu^2}{c T K_2\left(\frac{mc^2}{T}\right)}
\exp\left(-\frac{mc^2}{T}\sqrt{1+\frac{u^2}{c^2}}\right),
\end{equation}
or in terms of the Lorentz factor $\gamma$ as 
\begin{equation}
\label{eq:juttner-g}
g_\gamma\left(\gamma\right) = \frac{mc^2}{T}\frac{\gamma\sqrt{\gamma^2-1}}{K_2\left(\frac{mc^2}{T}\right)}
\exp\left(-\frac{mc^2}{T}\gamma\right), 
\end{equation}
where 
\begin{equation*}
\int_{1}^{\infty} g_{\gamma}\left(\gamma\right) {\rm d}\gamma = 1. 
\end{equation*}
This integral is confirmed with Mathematica. 

Similarly to Eq. (\ref{eq:rela-d-maxwellian}), the shifted Maxwell-J\"{u}ttner distribution 
in terms of the momentum vector is given with the boosted Lorentz factor $\gamma_B$ 
as \citep{Ueno_2021}
\begin{equation}
\label{eq:juttner-d}
g\left(\Vec{u}\right) = \frac{m}{4\pi c\gamma_D T K_2\left(\frac{mc^2}{T}\right)}
\exp\left\{-\frac{mc^2}{T}\gamma_B\left(\Vec{u}\right)\right\}.
\end{equation}
This is rewritten in terms of the boosted Lorentz factor as follows, 
\begin{equation}
\label{eq:juttner-d-g}
g_{\gamma_B}\left(\gamma_B\right) = \frac{mc^2}{TK_2\left(\frac{mc^2}{T}\right)}
\gamma_B\sqrt{\gamma_B^2-1}
\exp\left(-\frac{mc^2}{T}\gamma_B\right).
\end{equation}

With the normalized energy ${\cal E}' = mc^2(\gamma_B-1)/T$, 
equation (\ref{eq:juttner-d-g}) is rewritten as 
\begin{equation}
\label{eq:juttner-d-e}
g_{\cal E} \left({\cal E}',T \right) = \frac{T^2}{m^2c^4K_2\left(\frac{mc^2}{T}\right)}
\sqrt{{\cal E}'\left({\cal E}'+2\frac{mc^2}{T}\right)} 
\left({\cal E}'+\frac{mc^2}{T}\right)
\exp\left\{-\left({\cal E}'+\frac{mc^2}{T}\right)\right\}, 
\end{equation}
where
\begin{equation*}
\int_{0}^{\infty} g_{\cal E} \left({\cal E}' \right) {\rm d}{\cal E}' = 1. 
\end{equation*}
This integral is confirmed with Mathematica.
There is no analytic expression of the cumulative distribution of Eq. (\ref{eq:juttner-d-e}).

\begin{figure}[t]
\center
\includegraphics[width=0.5\textwidth,bb=0 0 640 580]{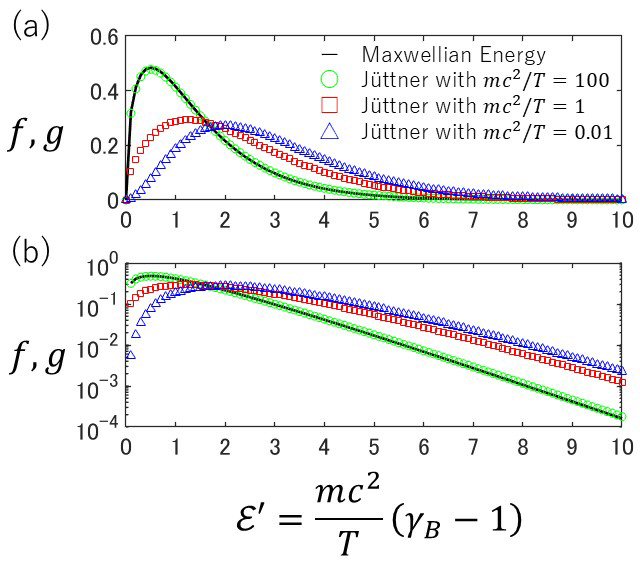}
\caption{
Comparison between the Maxwellian energy distribution 
in Eq. (\ref{eq:rela-d-maxwellian-e}) and 
the Maxwell-J\"{u}ttner distribution in Eq. (\ref{eq:juttner-d-e}) 
with different $mc^2/T$ in (a) a linear scale and (b) a logarithmic scale. 
The solid line shows the profile of Eq. (\ref{eq:rela-d-maxwellian-e}) with $\gamma_D=1$. 
The circles show the profile of Eq. (\ref{eq:juttner-d-e}) with $mc^2/T=100$. 
The squares show the profile of Eq. (\ref{eq:juttner-d-e}) with $mc^2/T=1$. 
The triangles show the profile of Eq. (\ref{eq:juttner-d-e}) with $mc^2/T=0.01$. 
}
\label{fig:4}
\end{figure}

Figure \ref{fig:4} shows the comparison between 
the relativistic Maxwellian energy distribution in Eq. (\ref{eq:rela-d-maxwellian-e}) 
and the Maxwell-J\"{u}ttner distribution in Eq. (\ref{eq:juttner-d-e}) 
as a function of the normalized energy ${\cal E}'$. 
The solid line shows the profile of Eq. (\ref{eq:rela-d-maxwellian-e}) with $\gamma_D=1$. 
The circles show the profile of Eq. (\ref{eq:juttner-d-e}) with $mc^2/T=100$. 
The squares show the profile of Eq. (\ref{eq:juttner-d-e}) with $mc^2/T=1$. 
The triangles show the profile of Eq. (\ref{eq:juttner-d-e}) with $mc^2/T=0.01$. 
With a smaller $T$, the Maxwell-J\"{u}ttner distribution approaches to 
the Maxwellian energy distribution. 
However, the high-energy tail of the Maxwell-J\"{u}ttner distribution is more enhanced 
with a larger $T$. 
The Maxwell-J\"{u}ttner distribution is expressed as a function of both ${\cal E}'$ and $T$. 
It is not easy to approximate the cumulative distribution of Eq. (\ref{eq:juttner-d-e}) 
by using an invertible function such as Eq. (\ref{eq:rela-d-maxwellian-e-app}), 
because the coefficients are given as a function of $T$. 

Alternatively, rejection sampling was applied to the reduced Maxwell-J\"{u}ttner distribution \citep{Swisdak_2013,Zenitani_2024}, 
in which the Maxwell-J\"{u}ttner distribution in Eq. (\ref{eq:juttner-d}) 
is transforming to cylindrical coordinates and 
analytically integrated over the perpendicular momentum space as follows 
\citep{Swisdak_2013,Melzani_2013}, 
\begin{equation}
\label{eq:juttner-d-ux}
g_{u_x}\left(u_x\right) = \frac{T}{2mc^3\gamma_D^3K_2\left(\frac{mc^2}{T}\right)}
\left(1+\frac{mc^2\gamma_D}{T}\sqrt{1+\frac{u_x^2}{c^2}}\right)
\exp\left\{-\frac{mc^2\gamma_D}{T}\left(\sqrt{1+\frac{u_x^2}{c^2}}-\frac{v_Du_x}{c^2}\right)\right\}.
\end{equation}
This is reconfirmed by using Mathematica as well as a numerical integration.

\begin{figure}[t]
\center
\includegraphics[width=1.0\textwidth,bb=0 0 1280 520]{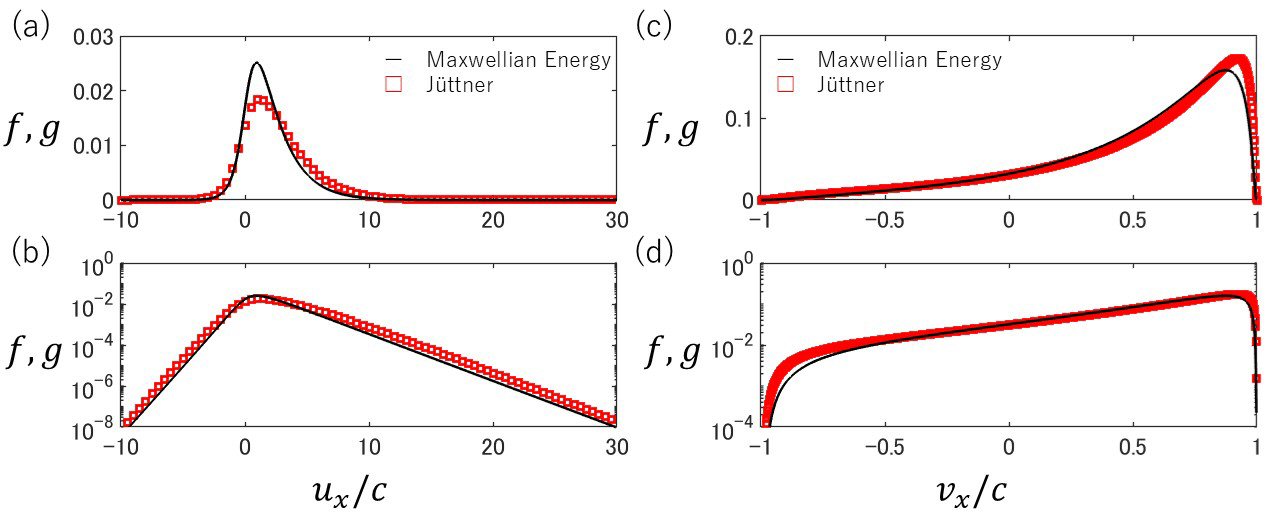}
\caption{
Comparison between tbetween 
the relativistic Maxwellian energy distribution in Eq. (\ref{eq:rela-d-maxwellian}) 
and the Maxwell-J\"{u}ttner distribution in Eq. (\ref{eq:juttner-d}) 
with $v_D/c=0.5$ and $mc^2/T=1$. 
Left panels show the reduced momentum distribution in the $u_x$ space 
in (a) a linear scale and (b) a logarithmic scale. 
Right panels show the reduced velocity distribution in the $v_x$ space 
in (c) a linear scale and (d) a logarithmic scale. 
The solid line shows the profile of the reduced Maxwellian energy distribution. 
The squares show the profile of the reduced Maxwell-J\"{u}ttner distribution. 
}
\label{fig:5}
\end{figure}

Figure \ref{fig:5} shows the comparison between 
the relativistic Maxwellian energy distribution in Eq. (\ref{eq:rela-d-maxwellian}) 
and the Maxwell-J\"{u}ttner distribution in Eq. (\ref{eq:juttner-d}) 
with $v_D/c=0.5$ and $mc^2/T=1$. 
Left panels show the reduced momentum distribution in the $u_x$ space 
in (a) a linear scale and (b) a logarithmic scale. 
Right panels show the reduced velocity distribution in the $v_x$ space 
in (c) a linear scale and (d) a logarithmic scale. 
The solid lines show the reduced Maxwellian energy distribution as a function 
of the momentum $u_x$ and the velocity $v_x$,  
which are obtained by using a numerical integration 
over the $u_y-u_z$ space and $v_y-v_z$ space, respectively. 
The squares show the reduced Maxwell-J\"{u}ttner distribution in Eq. (\ref{eq:juttner-d}) 
obtained by a numerical integration. 
Note that the reduced Maxwell-J\"{u}ttner distribution in the $u_x$ space is given 
in Eq. (\ref{eq:juttner-d-ux}) as well.

With a larger $mc^2/T$ (i.e., smaller $T$), the Maxwell-J\"{u}ttner distribution 
approaches to the Maxwellian energy distribution. 
With a smaller $mc^2/T$ (i.e., larger $T$), high-energy components are more enhanced 
in the Maxwell-J\"{u}ttner distribution, and the peak of the distribution shifts toward a high energy side. 

In both velocity distributions, the mean velocity is given as follows, 
\begin{equation}
\int_{-c}^{c} \int_{-c}^{c} \int_{-c}^{c} 
\Vec{v} \gamma^5(\Vec{v}) f\left(\frac{c\Vec{v}}{\sqrt{c^2-\left|\Vec{v}\right|^2}}\right) {\rm d}^3\Vec{v} 
= 
\int_{-c}^{c} \int_{-c}^{c} \int_{-c}^{c} 
\Vec{v} \gamma^5(\Vec{v}) g\left(\frac{c\Vec{v}}{\sqrt{c^2-\left|\Vec{v}\right|^2}}\right) {\rm d}^3\Vec{v} 
= \Vec{v}_D, 
\end{equation}
which is confirmed by a numerical integration. 

The mean momentum of the Maxwell-J\"{u}ttner distribution is given as follows \citep{Ueno_2021}, 
\begin{equation}
\label{eq:momentum_MJ}
\int_{-\infty}^{\infty} \int_{-\infty}^{\infty} \int_{-\infty}^{\infty} 
\Vec{u} g\left(\Vec{u}\right) {\rm d}^3\Vec{u} 
= \frac{K_3\left(\frac{mc^2}{T}\right)}{K_2\left(\frac{mc^2}{T}\right)}\gamma_D \Vec{v}_D, 
\end{equation}
which is reconfirmed by a numerical integration. 
However, the mean momentum of the Maxwellian energy distribution is given as follows 
(see \ref{sec:property} for detail), 
\begin{equation*}
\int_{-\infty}^{\infty} \int_{-\infty}^{\infty} \int_{-\infty}^{\infty} 
\Vec{u} f\left(\Vec{u}\right) {\rm d}^3\Vec{u} 
\end{equation*}
\begin{equation}
\label{eq:momentum_Maxwell}
= \left( \frac{4}{3} + \frac{2\gamma_DT}{mc^2} 
- \frac{2mc^2}{3\gamma_DT} \left[ 1-\sqrt{\frac{\pi mc^2}{\gamma_DT}} \exp\left(\frac{mc^2}{\gamma_DT}\right)
\left\{ 1- {\rm erf}\left(\sqrt{\frac{mc^2}{\gamma_DT}}\right) \right\}  
\right] 
\right) \gamma_Dv_D, 
\end{equation}
which is slightly smaller than the mean momentum of 
the Maxwell-J\"{u}ttner distribution in Eq. (\ref{eq:momentum_MJ}).

A useful property is obtained 
by performing the following integral for the Maxwell-J\"{u}ttner distribution, 
\begin{equation}
\label{eq:thermal_MJ}
\int_{-\infty}^{\infty} \int_{-\infty}^{\infty} \int_{-\infty}^{\infty} 
mc^2\left\{\gamma_B\left(\Vec{u}\right)-1\right\} g\left(\Vec{u}\right) {\rm d}^3\Vec{u} 
= 
T\left[ 3 - \frac{mc^2}{T}\left\{ 1-\frac{K_1\left(\frac{mc^2}{T}\right)}{K_2\left(\frac{mc^2}{T}\right)} \right\}
\right]. 
\end{equation}
This integral is confirmed with Mathematica.
This property is related to the thermal energy, which approaches to $3T/2$ 
with larger $mc^2/T$ but is enhanced with smaller $mc^2/T$. 
The property related to the thermal energy for the Maxwellian energy distribution is given as follows, 
\begin{equation}
\label{eq:thermal_Maxwell}
\int_{-\infty}^{\infty} \int_{-\infty}^{\infty} \int_{-\infty}^{\infty} 
mc^2\left\{\gamma_B\left(\Vec{u}\right)-1\right\} f\left(\Vec{u}\right) {\rm d}^3\Vec{u} 
= \frac{3}{2}\gamma_DT. 
\end{equation}

The kinetic energy of the Maxwell-J\"{u}ttner distribution is obtained by performing the following integral, 
\begin{equation*}
\int_{-\infty}^{\infty} \int_{-\infty}^{\infty} \int_{-\infty}^{\infty} 
mc^2\left\{\gamma\left(\Vec{u}\right)-1\right\} g\left(\Vec{u}\right) {\rm d}^3\Vec{u} 
\end{equation*}
\begin{equation}
\label{eq:kinetic_MJ}
= \frac{T}{\gamma_D} \left[ 3 - \frac{mc^2}{T} 
\left\{ 1 - \frac{K_1\left(\frac{mc^2}{T}\right)}{K_2\left(\frac{mc^2}{T}\right)} \right\}
\right]
+ \left\{ \frac{K_3\left(\frac{mc^2}{T}\right)}{K_2\left(\frac{mc^2}{T}\right)} - \frac{1}{\gamma_D+1}
\right\} m \gamma_D v^2_D. 
\end{equation}
This integral is confirmed with an numerical integration.
The kinetic energy of the Maxwellian energy distribution is given as follows
(see \ref{sec:property} for detail), 
\begin{equation*}
\int_{-\infty}^{\infty} \int_{-\infty}^{\infty} \int_{-\infty}^{\infty} 
mc^2\left\{\gamma_B\left(\Vec{u}\right)-1\right\} f\left(\Vec{u}\right) {\rm d}^3\Vec{u} 
= \frac{3}{2}T +
\end{equation*}
\begin{equation}
\left[ \frac{4}{3} -\frac{1}{\gamma_D+1} + \frac{2\gamma_DT}{mc^2} 
- \frac{2mc^2}{3\gamma_DT} \left\{ 1 - \sqrt{\frac{\pi mc^2}{\gamma_DT}} \exp\left(\frac{mc^2}{\gamma_DT}\right)
{\rm erfc}\left(\sqrt{\frac{mc^2}{\gamma_DT}}\right) \right\} 
\right] m \gamma_D v^2_D. 
\end{equation}
This integral is confirmed with an numerical integration.
The kinetic energy of the Maxwellian energy distribution is smaller than 
that of the Maxwell-J\"{u}ttner distribution in Eq. (\ref{eq:kinetic_MJ}).

In both distributions, the bulk drift energy is given as follows, 
\begin{equation}
{\cal D} = m\Vec{v}_D \cdot \left( \langle\Vec{u}\rangle - \frac{\gamma_D \Vec{v}_D}{\gamma_D+1} \right), 
\end{equation}
where $\langle\Vec{u}\rangle$ denotes the mean momentum vector. 
Note that the mean velocity vector is given as $\langle\Vec{v}\rangle = \Vec{v}_D$ as described above. 
If $\langle\Vec{u}\rangle = \gamma_D\Vec{v}_D$, the bulk drift energy is expressed as 
$\left|\gamma_D\Vec{v}_D\right|^2/(\gamma_D+1)$. 
In a non-relativistic limit (i.e., $\gamma_D \rightarrow 1$), 
the drift energy approaches to $m\left|\Vec{v}_D\right|^2/2$. 

It is able to separate the kinetic energy into the bulk drift energy and the thermal energy. 
The thermal energy of the Maxwellian energy distribution is given as ${\cal T} = 3T/2$. 
However, the thermal energy of the Maxwell-J\"{u}ttner distribution includes 
the enhancement due to $mc^2/T$ 
and the decrease by a factor of $\gamma_D$ due to the definition of the distribution.

\section{Summary}

Numerical methods for generating random variates from a relativistic Maxwellian-type distribution 
are important in relativistic particle kinetic simulations. 
Conventionally, rejection sampling is widely used 
to generate random variates of momentum vector from the Maxwell-J\"{u}ttner distribution. 
Inverse transforms sampling has issues in integrability of a target distribution and 
invertibility of its cumulative distribution. 

In the present study, a relativistic shifted-Maxwellian energy distribution is introduced 
as an alternative to the Maxwell-J\"{u}ttner distribution. 
To adopt the inverse transforms sampling, 
an invertible function is presented, which accurately approximate 
the cumulative distribution of the Maxwellian energy distribution. 
Then, random variates of momentum vector from the relativistic shifted-Maxwellian energy distribution 
are generated by utilizing three sets of random variables from uniform distribution. 
The numerical procedure of the present method is simple and 
easily implemented as an elemental function, 
i.e., $(u_x,u_y,u_z)= {\rm func}\left\{ R^{(1)},R^{(2)},R^{(3)} \right\}$, 
which is more suitable for high-performance computing.

\appendix

\section{Jacobian Determinant}
\label{sec:determinant}

From Eq. (\ref{eq:rela-d-maxwellian-rand}), the Jacobian matrix is given as 
\begin{equation*}
\mathbf{J} = \left[
\begin{array}{ccc}
\frac{\partial u_x}{\partial \gamma_B} & \frac{\partial u_x}{\partial \theta} & \frac{\partial u_x}{\partial \phi} \\
\frac{\partial u_y}{\partial \gamma_B} & \frac{\partial u_y}{\partial \theta} & \frac{\partial u_y}{\partial \phi} \\
\frac{\partial u_z}{\partial \gamma_B} & \frac{\partial u_z}{\partial \theta} & \frac{\partial u_z}{\partial \phi} \\
\end{array} \right]
\end{equation*}
\begin{equation}
= \left[
\begin{array}{ccc}
\gamma_D\left(\frac{c\gamma_B}{\sqrt{\gamma_B^2-1}}\cos\theta+v_D\right) & -c\gamma_D\sqrt{\gamma_B^2-1}\sin\theta & 0 \\
\frac{c\gamma_B}{\sqrt{\gamma_B^2-1}}\sin\theta\cos\phi & c\sqrt{\gamma_B^2-1}\cos\theta\cos\phi & -c\sqrt{\gamma_B^2-1}\sin\theta\sin\phi \\
\frac{c\gamma_B}{\sqrt{\gamma_B^2-1}}\sin\theta\sin\phi & c\sqrt{\gamma_B^2-1}\cos\theta\sin\phi & c\sqrt{\gamma_B^2-1}\sin\theta\cos\phi \\
\end{array} \right].
\end{equation}
Then, the Jacobian determinant is given as 
\begin{equation}
\label{eq:det}
{\rm det}\left(\mathbf{J}\right) = 
c^3\gamma_D\gamma_B\sqrt{\gamma_B^2-1}\sin\theta+
c^2v_D\gamma_D(\gamma_B^2-1)\cos\theta\sin\theta.
\end{equation}
This leads to Eq. (\ref{eq:convert_gb}). 

From Eqs. (\ref{eq:gmb}) and (\ref{eq:rela-d-maxwellian-rand}a), 
equation (\ref{eq:det}) is rewritten as follows, 
\begin{equation*}
{\rm det}\left(\mathbf{J}\right) = 
c^3\gamma_D\sqrt{\gamma_B^2-1}\sin\theta
\left(\gamma_B+\frac{v_Du_{x}}{c^2\gamma_D} - \frac{\gamma_{B} v_D^2}{c^2} \right)
=
c^3\sqrt{\gamma_B^2-1}\sin\theta
\left(\frac{\gamma_B}{\gamma_D}+\frac{v_Du_{x}}{c^2} \right)
\end{equation*}
\begin{equation}
=
c^3\gamma\sqrt{\gamma_B^2-1}\sin\theta.
\end{equation}

\section{Properties of Maxwellian Energy Distribution}
\label{sec:property}

Analytic expression of the mean momentum is obtained by using Eqs. (\ref{eq:rela-d-maxwellian-rand}a) 
and (\ref{eq:rela-d-maxwellian-gt}), 
\begin{equation*}
\int_{-\infty}^{\infty} \int_{-\infty}^{\infty} \int_{-\infty}^{\infty} 
u_x f\left(\Vec{u}\right) {\rm d}^3\Vec{u} 
= \int_0^\pi \int_1^\infty
\frac{1}{\sqrt{\pi}}
\left(\sqrt{\frac{mc^2}{\gamma_DT}}\right)^3
\gamma_D\left( c\sqrt{\gamma_{B}^2-1} \cos\theta + \gamma_{B} v_D \right)
\end{equation*}
\begin{equation*}
\left(
1+\frac{v_D}{c\gamma_B}\sqrt{\gamma_B^2-1}\cos\theta
\right)
\sqrt{\gamma_B-1}\sin\theta
\exp\left\{-\frac{mc^2}{\gamma_DT}\left(
\gamma_B
-1\right)\right\}
{\rm d}\gamma_B{\rm d}\theta
\end{equation*}
%
\begin{equation*}
= \frac{2\gamma_Dv_D}{\sqrt{\pi}}
\left(\sqrt{\frac{mc^2}{\gamma_DT}}\right)^3
\int_1^\infty
\left( \frac{1}{3} \frac{\gamma_B^2-1}{\gamma_B} 
+ \gamma_{B} \right)
\sqrt{\gamma_B-1}
\exp\left\{-\frac{mc^2}{\gamma_DT}\left(
\gamma_B
-1\right)\right\}
{\rm d}\gamma_B
\end{equation*}
%
\begin{equation*}
=\left[ \frac{4}{3} + \frac{2\gamma_DT}{mc^2} 
- \frac{2mc^2}{3\gamma_DT} \left\{ 1 - \sqrt{\frac{\pi mc^2}{\gamma_DT}} \exp\left(\frac{mc^2}{\gamma_DT}\right)
{\rm erfc}\left(\sqrt{\frac{mc^2}{\gamma_DT}}\right) \right\} 
\right] \gamma_Dv_D, 
\end{equation*}
where ${\rm erfc}(x)=1-{\rm erf}(x)$ is the complementary error function. 
This is confirmed with Mathematica.

Analytic expression of the kinetic energy is obtained by using Eqs. (\ref{eq:momentum_Maxwell}) 
and (\ref{eq:thermal_Maxwell}), 
\begin{equation*}
\int_{-\infty}^{\infty} \int_{-\infty}^{\infty} \int_{-\infty}^{\infty} 
mc^2\left\{\gamma\left(\Vec{u}\right)-1\right\} f\left(\Vec{u}\right) {\rm d}^3\Vec{u} 
\end{equation*}
\begin{equation*}
=\frac{1}{\gamma_D} \int_{-\infty}^{\infty} \int_{-\infty}^{\infty} \int_{-\infty}^{\infty} 
mc^2\left\{ \gamma_B\left(\Vec{u}\right) - 1\right\} f\left(\Vec{u}\right) {\rm d}^3\Vec{u} 
-mc^2
\end{equation*}
\begin{equation*}
+\frac{1}{\gamma_D} \int_{-\infty}^{\infty} \int_{-\infty}^{\infty} \int_{-\infty}^{\infty} 
mc^2\left( \frac{v_Du_x}{c^2} + 1 \right) f\left(\Vec{u}\right) {\rm d}^3\Vec{u} 
\end{equation*}
\begin{equation*}
= \frac{3T}{2} 
+ \frac{mc^2}{\gamma_D}\left(1-\gamma_D\right)
+ \left[ \frac{4}{3} + \frac{2\gamma_DT}{mc^2} 
- \frac{2mc^2}{3\gamma_DT} \left\{ 1 - \sqrt{\frac{\pi mc^2}{\gamma_DT}} \exp\left(\frac{mc^2}{\gamma_DT}\right)
{\rm erfc}\left(\sqrt{\frac{mc^2}{\gamma_DT}}\right) \right\} 
\right] m\gamma_D v^2_D 
\end{equation*}
\begin{equation*}
= \frac{3T}{2} 
+ \left[ \frac{4}{3} -\frac{1}{\gamma_D+1} + \frac{2\gamma_DT}{mc^2} 
- \frac{2mc^2}{3\gamma_DT} \left\{ 1 - \sqrt{\frac{\pi mc^2}{\gamma_DT}} \exp\left(\frac{mc^2}{\gamma_DT}\right)
{\rm erfc}\left(\sqrt{\frac{mc^2}{\gamma_DT}}\right) \right\} 
\right] m \gamma_D v^2_D. 
\end{equation*}

\section{Loading of Anisotropic Distribution}

Anisotropic Maxwell-J\"{u}ttner distributions are presented 
in Refs. \citep{Livadiotis_2016,Treumann_2016}. 
A shifted and anisotropic Maxwellian energy distribution is introduced 
as an extension of Ref. \citep{Treumann_2016}, 
\begin{subequations}
\begin{equation}
\label{eq:gma}
\gamma_A = \gamma_D \left( \sqrt{1 + \frac{T_\perp}{T_\parallel}\frac{u_x^2}{c^2} 
+ \frac{u_y^2}{c^2} + \frac{u_z^2}{c^2} } - \frac{v_Du_x}{c^2} \right),
\end{equation}
\begin{equation}
\label{eq:rela-d-maxwellian-a}
f\left(\Vec{u}\right) = \frac{1}{2\gamma_D} 
\left(\sqrt{\frac{m}{\pi \gamma_D T_\perp}}\right)^3
\frac{1}{
\gamma_A\left(\Vec{u}\right)
\sqrt{
\gamma_A\left(\Vec{u}\right)
+1} }
\exp\left[-\frac{mc^2}{\gamma_D T_\perp}\left\{
\gamma_A\left(\Vec{u}\right)
-1\right\}\right]. 
\end{equation}
\end{subequations}
Here, the momentum vector coordinate is taken so that 
the $u_x$ axis is parallel to the drift velocity vector $\Vec{v}_D$, and 
$T_\parallel$ and $T_\perp$ are temperature components 
parallel and perpendicular to the drift velocity vector, respectively. 

In addition to the following two conditions for ${\rm d}\gamma_B/{\rm d}t=0$, 
$\Vec{v}_D \cdot \Vec{E} = 0$ and $\Vec{E} \cdot \Vec{B}=0$, 
the third condition, $\Vec{E}=0$, is necessary to satisfy ${\rm d}\gamma_A/{\rm d}t=0$. 
That is, the drift velocity vector much be parallel to the magnetic field, $\Vec{v}_D\times\Vec{B}=0$ 
in relativistic shifted and anisotropic distributions. 

Random of energy variates from Maxwellian energy distribution are generated 
by using Eq. (\ref{eq:rela-d-maxwellian-e-rand}). 
The energy variates are converted to the Lorentz factor variates as follows,
\begin{equation}
\gamma_{A,n} = 1 + \frac{\gamma_DT_\perp}{mc^2}{\cal E}_n.
\end{equation}
Then, the random variates of momentum vector are obtained as follows, 
\begin{subequations}
\begin{equation}
u_{x,n} = \gamma_D \sqrt{\frac{T_\parallel}{T_\perp}}\left( c\sqrt{\gamma_{A,n}^2-1} \cos\theta_n + \gamma_Av_D \right), 
\end{equation}
\begin{equation}
u_{y,n} = c\sqrt{\gamma_{A,n}^2-1} \sin\theta_n \cos\phi_n,
\end{equation}
\begin{equation}
u_{z,n} = c\sqrt{\gamma_{A,n}^2-1} \sin\theta_n \sin\phi_n, 
\end{equation}
\end{subequations}
where $\theta_n$ and $\phi_n$ are given by using Eq. (\ref{eq:summary}b,c)
and Eq. (\ref{eq:rela-d-maxwellian-gt-cum-inv}) by replacing $\gamma_B$ with $\gamma_A$.

\section{Loading of Non-relativistic Maxwellian Distribution}

Random of energy variates from Maxwellian energy distribution are generated 
by using Eq. (\ref{eq:rela-d-maxwellian-e-rand}). 
The energy variates are converted to scalar velocity variates as follows,
\begin{equation}
v_n = \sqrt{\frac{2T}{m}{\cal E}_n}.
\end{equation}
Then, the random variates of velocity vector are obtained as follows, 
\begin{subequations}
\begin{equation}
v_{x,n} = v_n \cos\theta_n + v_D, 
\end{equation}
\begin{equation}
v_{y,n} = v_n \sin\theta_n \cos\phi_n,
\end{equation}
\begin{equation}
v_{z,n} = v_n \sin\theta_n \sin\phi_n, 
\end{equation}
\end{subequations}
where $\theta_n$ and $\phi_n$ are given by using Eq. (\ref{eq:theta-phi}).


\section*{Code availability}
A sample code 
for generating a relativistic shifted-Maxwellian energy distribution 
in MATLAB \citep{Matlab} 
is available from \url{https://github.com/taka-umeda/rela.maxwellian}.


\section*{Declaration of generative AI and AI-assisted technologies in the writing process}
During the preparation of the final manuscript, 
the author used Google Gemini for spelling and grammar corrections. 
After using this tool, the author reviewed and edited the content as needed 
and take full responsibility for the content of the publication.


\section*{Acknowledgment}
This work was supported by MEXT/JSPS under 
Grant-in-Aid (KAKENHI) for Scientific Research (B) No.JP24K00603.  
A part of numerical calculations were performed on the Interdisciplinary Large-scale Computing System 
at Information Initiative Center, Hokkaido University. 
This work was also conducted under joint research programs 
at Institute for Space-Earth Environmental Research, Nagoya University.

\end{document}